\documentclass[prb,twocolumn,superscriptaddress,showpacs,showkeys]{revtex4}
\usepackage[normalem]{ulem}

\usepackage{graphics}% Include figure files
\usepackage{color}
\usepackage{amsmath}% for advanced Eq. numbering
\usepackage{amssymb}% for advanced Eq. numbering
\usepackage[colorlinks=false]{hyperref}
\usepackage{pslatex}
\usepackage{amsfonts}
\usepackage{amsthm}
\usepackage{amscd}
\usepackage{graphicx}

\include{MyMc} %simple macros allowed

\begin{document}

\title{
  Ground states of one and two fractional vortices in long Josephson 0-$\kappa$-junctions
}

\author{E.~Goldobin}
\email{gold@uni-tuebingen.de}
\author{D.~Koelle}
\author{R.~Kleiner}
\affiliation{
  Physikalisches Institut --- Experimentalphysik II,
  Universit\"at T\"ubingen,
  Auf der Morgenstelle 14,
  D-72076 T\"ubingen, Germany
}

\pacs{
  74.50.+r,   %Proximity effects, weak links, tunneling phenomena,
              %and Josephson effect
  85.25.Cp    %Josephson devices
  74.20.Rp    %Pairing symmetries (other than s-wave)
}

\keywords{
  Long Josephson junction, sine-Gordon,
  half-integer flux quantum, semifluxon,
  0-pi-junction
}

%Acronims:
% LJJ:66, SF, PSF, NSF, AFM:20, ZFS:9

%\definecolor{gray}{gray}{0.75}
%\date{9.9.2002 [\colorbox{gray}{cond-mat/0209214}]}
\date{\today}

\begin{abstract}
  Half integer Josephson vortices in 0-$\pi$-junctions, discussed theoretically and observed experimentally, spontaneously appear at the point where the Josephson phase is $\pi$-discontinuous. The creation of \emph{arbitrary} discontinuities of the Josephson phase has been demonstrated recently.
  
  Here we study fractional vortices formed at an arbitrary $\kappa$-discontinuity, discuss their stability and possible ground states. The two stable states are not mirror symmetric. Furthermore, the possible ground states formed at two $\kappa$-discontinuities separated by a distance $a$ are investigated, and the energy and the regions of stability of each ground state are calculated. We also show that the ground states may strongly depend on the distance $a$ between the discontinuities. There is a crossover distance $a_c$ such that for $a<a_c$ and for $a>a_c$ the ground states may be qualitatively different.
\end{abstract}

%{\small
%  To be submitted to Phys.\ Rev.\ B (8+ pages)
%}

\maketitle

%\tableofcontents

\section{Introduction}
\label{Sec:Intro}

During the last years it was shown both theoretically\cite{Bulaevskii:0-pi-LJJ,Xu:SF-shape,Kuklov:1995:Current0piLJJ,Kogan:3CrystalVortices,Goldobin:SF-Shape} and experimentally\cite{Kirtley:SF:HTSGB,Kirtley:SF:T-dep,Sugimoto:TriCrystal:SF,Hilgenkamp:zigzag:SF} that one can create and study half integer Josephson vortices which carry only half of the magnetic flux quantum $\Phi_0$. To create such vortices one usually has to use a so-called long Josephson 0-$\pi$-junction (0-$\pi$-LJJ), \ie, a junction of which one part behaves as a 0-junction (positive critical current) and the other part as a $\pi$-junction (negative critical current). There are several approaches and technologies which can be used to fabricate such junctions. For example, one can use junctions based on superconductors with anisotropic order parameter\cite{Tsuei:Review,Smilde:ZigzagPRL,Kogan:3CrystalVortices}, junctions with ferromagnetic barrier\cite{Ryazanov:2001:SFS-PiJJ,Kontos:2002:SIFS-PiJJ}, or even conventional junctions with a pair of tiny injectors\cite{Goldobin:Art-0-pi,Malomed:2004:ALJJ:Ic(Iinj),Ustinov:2002:ALJJ:InsFluxon}.

In fact, all this types of Josephson junctions may be described by a model in which the Josephson phase is $\pi$-discontinuous at the 0-$\pi$-boundary.
In an infinite LJJ, the presence of a discontinuity results in an infinite energy, which of course cannot be allowed in nature. To save energy, the Josephson phase $\phi$ bends around the discontinuity (on the length scale of the Josephson penetration depth $\lambda_J$) so that $\phi(\pm\infty)=2\pi n$. This localizes the energy in a region of size $\sim\lambda_J$ around the discontinuity and creates a local magnetic field $\propto\phi_x(x)$ and currents $\propto\sin\phi(x)$ circulating around the discontinuity. Circulating currents create a vortex with the total magnetic flux equal to $\Phi_0/2$. 

Instead of specifying the flux carried by the vortex, we just denote it by the total continuous change of the phase on the interval from $x=-\infty$ to $x=+\infty$ (not including the discontinuity), \ie, a $\pi$-vortex is a semifluxon carrying $\Phi_0/2$, while a $2\pi$-vortex is the usual integer fluxon carrying $\Phi_0$.

Recently it was demonstrated that it is possible to create an \emph{arbitrary} $\kappa$-discontinuity\cite{Goldobin:Art-0-pi,Buzdin:2003:phi-LJJ}. The question, therefore, is: what kind of vortices can be formed at such an arbitrary $\kappa$-discontinuity? As we saw above, the main reason of the fractional vortex formation is to save energy (localize the energy around the phase discontinuity point) by compensating the phase jump at the discontinuity towards an integer number of $2\pi$. In the case of a $+\pi$-discontinuity, the vortex is either a $-\pi$-vortex with $\phi(+\infty)-\phi(-\infty)=0$ or a $+\pi$-vortex with $\phi(+\infty)-\phi(-\infty)=2\pi$. The $\pm\pi$-vortices are mirror symmetric, \ie, one carries $+\Phi_0/2$ with currents circulating clockwise, while the other carries $-\Phi_0/2$ with currents circulating counterclockwise.

We can already anticipate that in the case of an arbitrary $-\kappa$-discontinuity, at least two, generally \emph{not mirror symmetric} vortices can exist: a $+\kappa$-vortex with $\phi(+\infty)-\phi(-\infty)=0$ and a  $(\kappa-2\pi)$-vortex with $\phi(+\infty)-\phi(-\infty)=-2\pi$. Below we present an analysis of the possible vortex states pinned at one or two arbitrary discontinuities of the Josephson phase.

The paper is organized as follows. In Sec.~\ref{Sec:Model} we introduce the model for the case of arbitrary discontinuities. In section \ref{Sec:OneVortex} we derive an analytical expression for the phase of various vortices pinned at an arbitrary discontinuity. Then in Sec.~\ref{Sec:TwoVortices} we study numerically the ground state of a LJJ with two $\kappa$-discontinuities situated at a distance $a$ from each other. Section \ref{Sec:Conclusion} concludes this work.

\section{The model}
\label{Sec:Model}

The behavior of the Josephson phase $\phi(x)$ in a LJJ with discontinuities can be described by the following perturbed sine-Gordon equation\cite{Goldobin:SF-Shape}
\begin{equation}
  \phi_{xx}-\phi_{tt}-\sin(\phi) = \alpha\phi_t-\gamma(x)+\theta_{xx}(x)
  , \label{Eq:sG-phi}
\end{equation}
where the subscripts $x$ and $t$ denote a partial derivative with respect to coordinate $x$ and time $t$, respectively; $\theta(x)$ is a step function which is $\kappa$ discontinuous at the points $x=x_i$ ($i=1,\ldots,N$) of the junction and is constant everywhere else. We note here, that for a $\pi$-discontinuity the sign of the jump does not matter. A junction with a $+\pi,+\pi,\ldots,+\pi$ set of discontinuities physically behaves in the same way as, \eg, a junction with $+\pi,-\pi,+\pi,-\pi,\ldots,+\pi,-\pi$. In contrast, the sign of a $\kappa$-discontinuity \emph{does matter}. As we will see later, a junction with two $(-\kappa,-\kappa)$ discontinuities may have completely different ground states than a junction with $(+\kappa,-\kappa)$ discontinuities. If the discontinuity is created by injectors\cite{Goldobin:Art-0-pi}, the sign of discontinuity depends on the polarity of the injected currents.

Equation~(\ref{Eq:sG-phi}) is written in normalized units. The coordinate is normalized to the Josephson penetration depth $\lambda_J$, the time is normalized to the inverse plasma frequency $\omega_p^{-1}$, the bias current density $\gamma=j/j_c$ is normalized to the critical current density, and $\alpha=1/\sqrt{\beta_c}$ is the dimensionless damping parameter ($\beta_c$ is the McCumber-Stewart parameter).

Looking at Eq.~(\ref{Eq:sG-phi}) one can anticipate that since $\phi_{xx}$ and $\theta_{xx}$ are two additive terms, the solution $\phi(x)$ should be discontinuous at $x=x_i$. It is not very practical to deal with discontinuous functions such as $\phi(x)$, and with singular functions such as $\theta_{xx}(x)$. Therefore, it is convenient, following Ref.~\onlinecite{Goldobin:SF-Shape}, to introduce a new \emph{continuous} phase $\mu$: $\phi(x,t)=\mu(x,t)+\theta(x)$. Then the sine-Gordon equation reads
\begin{equation}
  \mu_{xx}-\mu_{tt}-\sin(\mu+\theta) = \alpha\mu_t-\gamma(x)
  .\label{Eq:sG-mu}
\end{equation}

\section{Vortex at a single $\kappa$-discontinuity}
\label{Sec:OneVortex}

For the investigation of the ground state of a system with one $\kappa$-discontinuity at $x_1=0$ we use the static version of Eq.~(\ref{Eq:sG-mu}) without bias current ($\gamma=0$)
\begin{equation}
  \mu_{xx}=\sin(\mu+\theta)
  ,\label{Eq:sG-mu-static}
\end{equation}
and with 
\begin{equation}
 \theta(x)=\left\{
 \begin{array}{rl}
   0,       &x<0;\\
   -\kappa, &x>0,
 \end{array}
 \right.
 = -\kappa\Heavyside(x),
 \label{Eq:theta}
\end{equation}
where $\Heavyside(x)$ is a Heavyside step function.

The results in the case of a $+\kappa$ discontinuity are mirror symmetric, \ie, $\mu(x,\kappa)=-\mu(x,-\kappa)$. Since the phase is defined modulo $2\pi$, without loosing generality, below we consider only $0\leq\kappa\leq2\pi$.

The natural boundary conditions (BCs) at $x=0$ are
\begin{subequations}
  \begin{eqnarray}
   \mu  (+0) &=& \mu  (-0)
    ; \label{Eq:bc0_1}\\
   \mu_x(+0) &=& \mu_x(-0)
    . \label{Eq:bc0_2}
  \end{eqnarray}
  \label{Eq:bc-zero}
\end{subequations}

The BCs at $x\to\pm\infty$ come from the fact that the energy density of the system 
\begin{equation}
  {\cal H} = \frac12 \mu_x^2 + [1-\cos(\mu+\theta)]
  , \label{Eq:HamiltonianDensity}
\end{equation}
must vanish at $x\to\pm\infty$. This can be achieved when both $\phi_x\to0$ and $\phi\to 2\pi n$. Since $\phi_x(\pm\infty)=\mu_x(\pm\infty)$, $\phi(-\infty)=\mu(-\infty)$ and $\phi(+\infty)=\mu(+\infty)-\kappa$, without loosing generality we adopt the following BCs
\begin{subequations}
  \begin{eqnarray}
    \mu_x(\pm\infty)=0
    ; \label{Eq:bc-inf1}\\
    \mu(-\infty)=0
    ; \label{Eq:bc-inf2}\\
    \mu(+\infty)=\kappa+2\pi n
    . \label{Eq:bc-inf-charge}
  \end{eqnarray}
  \label{Eq:bc-inf}
\end{subequations}

First we consider a $+\kappa$-vortex, \eg, $n=0$ in Eq.~(\ref{Eq:bc-inf-charge}). The non-trivial solutions of Eq.~(\ref{Eq:sG-mu}) for $x<0$ and  $x>0$ are always given by the ``tails'' of a fluxon, which may be shifted along the $\mu$ or $x$ axes to satisfy the BCs\cite{Kogan:3CrystalVortices}, thus
\begin{equation}
  \mu(x) = \left\{
    \begin{array}{rl}
    \phi_0(x-x_0),        \quad& x<0,\\
    \kappa-\phi_0(-x-x_0),\quad& x>0,
    \end{array}\right.
  , \label{Eq:dvortex}
\end{equation}
where $\phi_0(x)$ is a soliton (fluxon) solution
\begin{equation}
  \phi_0(x)=4\arctan e^{x}
  . \label{Eq:phi_sol}
\end{equation}
Note, that the ansatz (\ref{Eq:dvortex}) already satisfies the BCs (\ref{Eq:bc-inf}). By trying to satisfy the BCs (\ref{Eq:bc-zero}) we arrive at the expression for $x_0$
\begin{equation}
  x_0 = -\ln\tan\frac\kappa8 > 0
  . \label{Eq:dvortex:x_0}
\end{equation}

Second, we consider a $(\kappa-2\pi)$-vortex, \eg, $n=-1$ in Eq.~(\ref{Eq:bc-inf-charge}). A similar procedure gives
\begin{equation}
  \mu(x) = \left\{
    \begin{array}{rl}
    \phi_0(x-x_0),\quad& x<0,\\
    \kappa-2\pi-\phi_0(-x-x_0),\quad& x>0,
    \end{array}\right.
  , \label{Eq:cvortex}
\end{equation}
with $\phi_0$ again given by Eq.~(\ref{Eq:phi_sol}) and 
\begin{equation}
  x_0 = -\ln\tan\left( \frac\kappa8-\frac\pi4 \right) >0
  . \label{Eq:cvortex:x_0}
\end{equation}

In principle one can try all other $n$ in Eq.~(\ref{Eq:bc-inf-charge}), but as soon as $|\kappa+2\pi n|>4\pi$ the solution cannot be constructed at all. Here, formally, the value of $x_0$, which in the general case is given by
\begin{equation}
  x_0 = - \ln\tan\left| \frac\kappa8+\frac{n\pi}4 \right|
  , \label{Eq:x_0:general}
\end{equation}
becomes complex. Physically this means that one cannot construct a localized energy solution out of two fluxon tails so that the phase changes by more than $4\pi$ when $x$ goes from $-\infty$ to $+\infty$. Moreover already for $|\kappa+2\pi n|>2\pi$ the constructed solution is unstable\cite{Goldobin:2KappaEigenModes}.

\begin{figure*}[!htb]
  \begin{center}
    \includegraphics{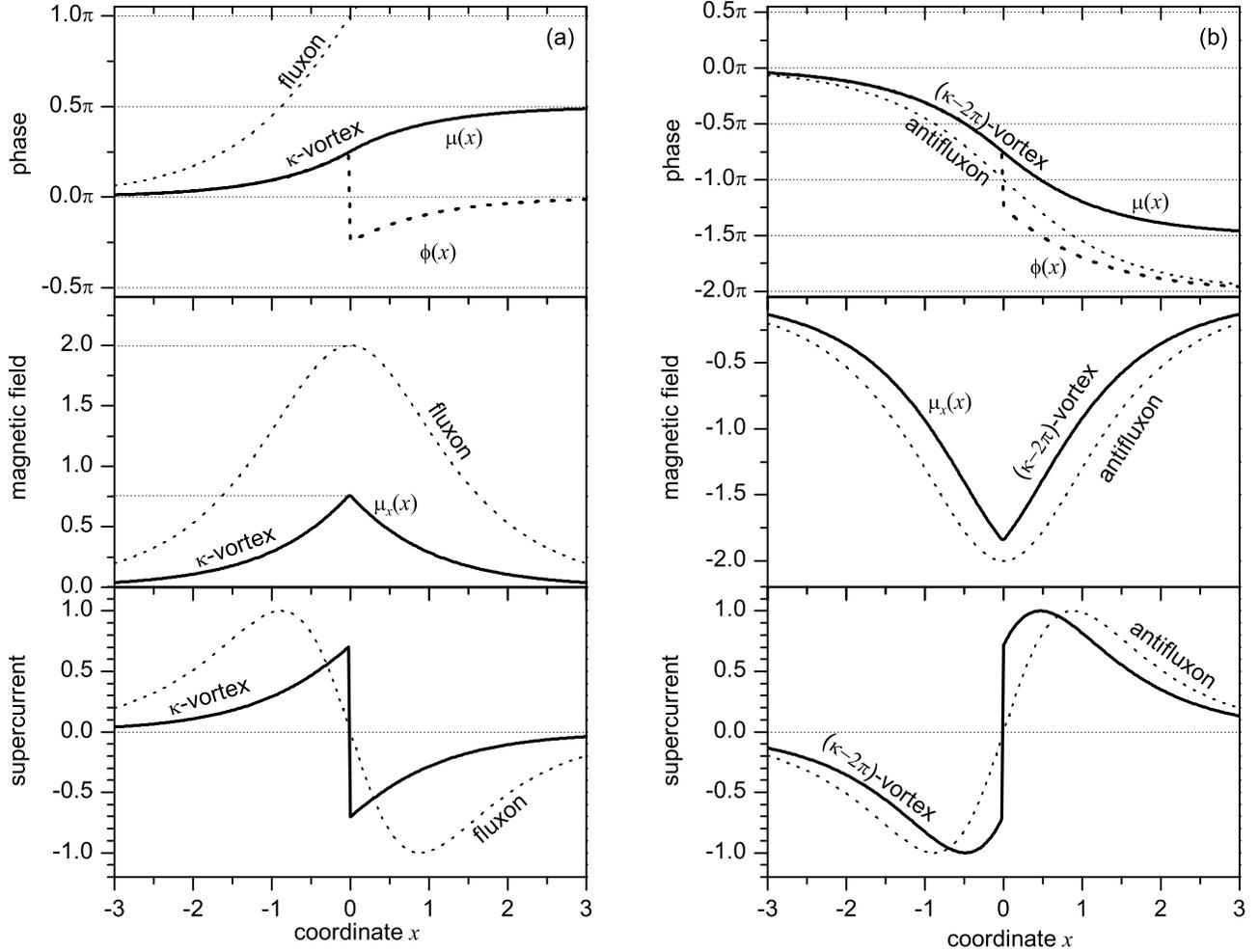}
  \end{center}  
  \caption{
    Phases $\phi(x)$ and $\mu(x)$, the magnetic field $\mu_x(x)$ and the supercurrent $\sin\phi(x)$ for $\kappa=\pi/2$. (a) direct $\pi/2$-vortex (\ref{Eq:dvortex}), (b) complementary $-3\pi/2$-vortex (\ref{Eq:cvortex}). The lines denoted as ``fluxon'' and ``antifluxon'' show the corresponding curves for $2\pi$-vortex (a) and $-2\pi$-vortex (b) for comparison.
  }
  \label{Fig:vortex}
\end{figure*}

Thus, for a $-\kappa$-discontinuity we have only two possible stable fractional vortices, given by Eqs.~(\ref{Eq:dvortex}) and (\ref{Eq:cvortex}), which can localize the energy around the discontinuity. In Fig.~\ref{Fig:vortex} we show both phases $\phi(x)$ and $\mu(x)$, the magnetic field $\mu_x(x)$ and the supercurrent $\sin\phi(x)$ for both of these solutions for the particular case of $\kappa=\pi/2$. Similar pictures for semifluxons ($\kappa=\pi$) can be seen in Ref.~\onlinecite{Goldobin:SF-Shape}.

In the following, a $\kappa$-vortex pinned at a $-\kappa$-discontinuity is called a \emph{direct} vortex and denoted as \state{u}, while a $(\kappa-2\pi)$-vortex pinned at a $-\kappa$-discontinuity is called a \emph{complementary} vortex and denoted as \state{D}. We intentionally avoid the word ``antivortex'' because when at $\kappa=2\pi$ the direct vortex is a fluxon, the complementary vortex is not an antifluxon. It is just a ``no vortex'' state, $\mu=2\pi n$. By definition, the complementary of a complementary vortex gives again a \emph{direct} $+\kappa$-vortex. In the general case there are only two stable vortices: a direct vortex and a complementary vortex. The single exception is the case of $\kappa=2\pi n$, for which there exist tree stable solutions: $+2\pi$-vortex (fluxon), constant  phase state (zero phase), and $-2\pi$-vortex (antifluxon). The complementary ``vortex'' for both $\pm2\pi$-vortices is a constant phase state and, vice versa, for the constant phase state the complementary vortex is either a fluxon or an antifluxon. In the majority of situations this can be distinguished due to the conservation of the topological charge. Note that in the notations such as \state{udUD} the direction of the arrows shows the polarity of the vortex (up or down), while the harpoon on the left or right side indicates whether the vortex is direct or complementary. The notations are summarized in Tab.~\ref{Tab:Notations}.

\begin{table}
  \begin{tabular}{cccl}
    \textbf{Symbol} & \textbf{Discontinuity} & 
    \textbf{Topological charge} & \textbf{Name}\\
    \hline
    \state{u} & $-\kappa$ & $+\kappa$ & direct\\
    \state{d} & $+\kappa$ & $-\kappa$ & direct\\
    \state{U} & $+\kappa$ & $2\pi-\kappa$ & complementary\\
    \state{D} & $-\kappa$ & $\kappa-2\pi$ & complementary\\
    \state{s} & $\pm\pi$ & $+\pi$ & semifluxon\\
    \state{a} & $\pm\pi$ & $-\pi$ & antisemifluxon\\
    \state{F} & $0$ & $+2\pi$ & fluxon\\
    \state{A} & $0$ & $-2\pi$ & antifluxon\\
    \hline
  \end{tabular}
  \caption{%
    Notations for different types of vortices.
  }
  \label{Tab:Notations}
\end{table}

The energy of the direct vortex can be easily calculated by integrating the energy density (\ref{Eq:HamiltonianDensity}) over the junction length with $\mu(x)$ given by (\ref{Eq:dvortex}), which gives
\begin{equation}
  E(\kappa) = 16 \sin^2\frac\kappa8
  . \label{Eq:E-kappa}
\end{equation}
The energy of a complementary vortex is obviously $E(2\pi-\kappa)$. Note, that this expression gives the right energy of a fluxon $E(2\pi)=8$ and of a semifluxon $E(\pi)=4(2-\sqrt2)$, see Ref.~\onlinecite{Goldobin:SF-ReArrange}.

One may use energy arguments to make some conclusions about the stability of various states\cite{Kogan:3CrystalVortices}. For example, can ``heavy'' vortex emit a fluxon and turn into a ``lighter'' complimentary vortex? If the energy of a fluxon together with the energy of a complimentary vortex is larger than the energy of an original vortex, this process is clearly impossible. Using Eq.~(\ref{Eq:E-kappa}) we can immediately see that such process is forbidden for vortices with the topological charge less than $2\pi$. On the other hand the $(\kappa+2\pi)$-vortex with the topological charge larger than $2\pi$ \emph{may} emit a fluxon and turn into a vortex with the topological charge $\kappa$. Indeed, using Eq.~(\ref{Eq:E-kappa}) one can prove that 
\[
  E(\kappa+2\pi)>E(2\pi)+E(\kappa)
  \mbox{ for\ \ }0<\kappa<2\pi.
\]
We stress that this is necessary, but not sufficient condition for vortex instability. Similarly, the $(\kappa-4\pi)$-vortex with the topological charge less than $-2\pi$, \emph{may} emit fluxon and turn itself into a $(\kappa-2\pi)$-vortex. More strict analysis\cite{Goldobin:2KappaEigenModes} shows that indeed the $\xi$-vortices with $|\xi|>2\pi$ 
%(the absolute value of the topological charge larger than $2\pi$) 
are unstable.

In comparison with a semifluxon, for which $\kappa$ and $(\kappa-2\pi)$ vortices are mirror symmetric, in the case of an arbitrary discontinuity the symmetry is broken. This may result in a number of interesting consequences both for ground states and for vortex dynamics.

We would like to note here that many authors in the context of unconventional superconductivity propose that the presence of a fractional vortex (not equal to $\Phi_0$ or $\Phi_0/2$) is the signature of time-reversal symmetry violation. In our case, the time-reversal symmetry is preserved: upon inversion of time a $\kappa$-discontinuity becomes a $-\kappa$-discontinuity, and the corresponding direct and complementary vortices change their signs too. In our case, only the parity is violated, \ie, direct and complementary vortices pinned at a fixed discontinuity are not mirror symmetric.

\section{Ground state of two vortices}
\label{Sec:TwoVortices}

\subsection{Possible states}

Now we consider two $\kappa$-discontinuities ($0<\kappa<2\pi$) in an infinite LJJ situated at $x_{1,2}=\pm a/2$ (at a distance $a$ from each other). If both discontinuities have the same sign of $\kappa$, \eg, ($-\kappa,-\kappa$) and 
\begin{equation}
  \theta(x) = -\kappa\left[ 
    \Heavyside\left( x+\frac{a}{2} \right) +
    \Heavyside\left( x-\frac{a}{2} \right)
  \right] + 2\pi n
  , \label{Eq:theta-mm}
\end{equation}
there are two possible irreducible vortex configurations: the symmetric ferromagnetic (FM) state $\state{uu}$ consisting of $(+\kappa,+\kappa)$-vortices, and the asymmetric antiferromagnetic (AFM) state $\state{uD}$ consisting of  $(+\kappa,\kappa-2\pi)$-vortices. If the discontinuities have different sign, \eg, $(+\kappa,-\kappa)$ and
\begin{equation}
  \theta(x) = \kappa\left[ 
    \Heavyside\left( x+\frac{a}{2} \right) -
    \Heavyside\left( x-\frac{a}{2} \right)
  \right] + 2\pi n
  , \label{Eq:theta-pm}
\end{equation}
there are two other irreducible vortex states: the asymmetric FM state $\state{Uu}$ consisting of $(2\pi-\kappa,\kappa)$-vortices, and the symmetric AFM state $\state{du}$ consisting of $(-\kappa,+\kappa)$-vortices. Below we consider these four ground states. 

The analytical description of the two vortex states is possible but involves rather bulky and intuitively not clear expressions with elliptic functions. Therefore, here we present only a qualitative picture of the ground states and calculate the necessary quantities numerically. For these numerical calculations we have used \textsc{StkJJ} software\cite{StkJJ}. To obtain the magnetic field profiles we started the simulations at $\kappa=\pi$ with the phase profile approximately corresponding to the one of four states. Then $\kappa$ was changed in small steps $\Delta{\kappa}=0.01\pi$ towards $0$ or towards $2\pi$. After changing $\kappa$ the program was waiting for the decay of all oscillations in the system and, after this, a snapshot of the magnetic field $\mu_x(x)$ was produced.

For an infinite LJJ containing more than one discontinuity and more than one vortex the topological charge of a vortex is not very well defined, especially if the vortices are close to each other. The topological charge of a vortex is given by $\mu(+\infty)-\mu(-\infty)$. This definition assumes that there are no other topological excitations in the vicinity of a vortex. When vortices are close to each other it may be difficult to separate them, their magnetic field and currents may overlap and superpose (cancel or enhance) non-linearly. Nevertheless we will still continue using the terms $\kappa$-vortex or $\kappa-2\pi$-vortex, especially in the case of weak coupling (large and moderate distance between vortices), assuming that the state is constructed from the corresponding single vortices at an infinite distance from each other and then the distance between discontinuities is smoothly and quasi-statically decreased down to desired values. We also assume that during this decrease no abrupt reconfigurations occurred in the system.

\begin{figure*}[!tb]
  \centering
  \includegraphics{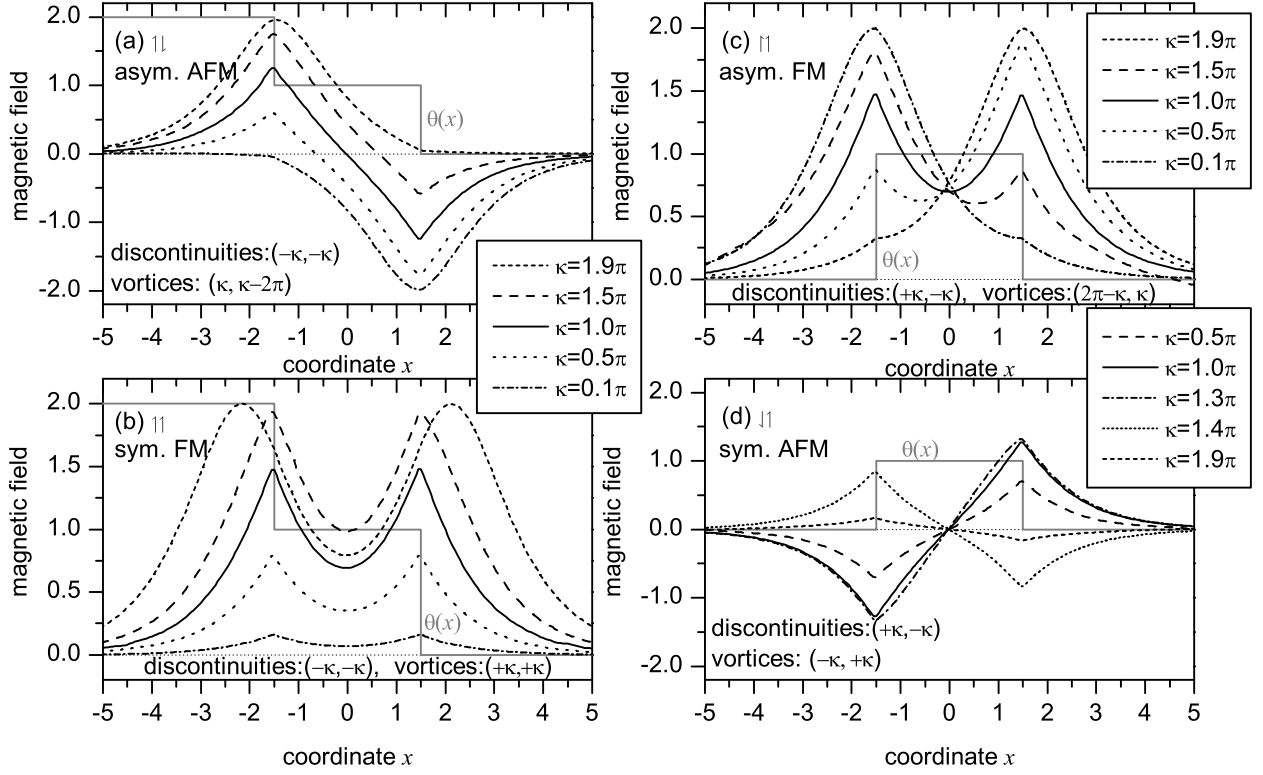}
  %\scalebox{0.9}{\includegraphics{disc-mm}\includegraphics{disc-pm}}
  \caption{
    Magnetic field profiles $\mu_x(x)$ for different values of $\kappa$ for the states (a) \protect\state{uD}, (b) \protect\state{uu}, (c) \protect\state{Uu} and (d) \protect\state{du} with discontinuities at $x=\pm a/2$ and $a=3$. The function $\theta(x)$ is shown by gray line. The amplitude of $\theta(x)$ is arbitrary.
  }
  \label{Fig:Vortex2:h}
\end{figure*}

First, we consider the case of a rather large $a=3$ (weak vortex-vortex interaction), for which the states can be constructed out of single vortex states. The value of $a$ is given in normalized units, \ie, in units of $\lambda_J$. The numerically calculated magnetic field profiles corresponding to the four different states 
\state{uD},  \state{uu}, \state{Uu} and \state{du} are shown in Fig.~\ref{Fig:Vortex2:h}. 

We introduce the term \emph{complementary configuration or state} which can be obtained from any original configuration if all vortices are substituted by the complementary ones. Note that the profiles $\mu_x^{c}$ of the complementary states 
\state{Du},  \state{DD}, \state{dD} and \state{UD} pinned at the same discontinuities are given by
\begin{equation}
  \mu_x^{c}(x,\kappa)=-\mu_x(x,2\pi-\kappa)
  . \label{Eq:mu_x-compl}
\end{equation}
If we discuss some property, \eg, instability, that happens with some state at some value of $\kappa$, the same thing happens with the complementary state at $2\pi-\kappa$. Thus all that is presented below can be easily mapped to the complementary states using $\kappa\to2\pi-\kappa$ and Eq.~(\ref{Eq:mu_x-compl}).

\paragraph{Asymmetric AFM state.} The field profiles $\mu_x(x)$ formed at the $(-\kappa,-\kappa)$ discontinuities and corresponding to the asymmetric AFM state \state{uD} are shown in Fig.~\ref{Fig:Vortex2:h}(a) for several values of $\kappa$. This state is stable for any value of $0<\kappa<2\pi$. At $\kappa=0$ or at $\kappa=2\pi$ the state becomes unstable, the integer fluxon is emitted and the state turns itself into a symmetric FM state. These processes can be written as
\[
  \state{uD}\stackrel{\kappa=0}{\longrightarrow}\state{uu}+\state{A}
  ,\mbox{ or }
  \state{uD}\stackrel{\kappa=2\pi}{\longrightarrow}\state{DD}+\state{F}.
\]

\paragraph{Symmetric FM state.} The field profiles $\mu_x(x)$ formed at the $(-\kappa,-\kappa)$ discontinuities and corresponding to the symmetric FM state \state{uu} are shown in Fig.~\ref{Fig:Vortex2:h}(b) for several values of $\kappa$. When the vortices become ``heavy'' their top gets rounded. This state is stable only for $0<\kappa<\kappa_c^{\state{uu}}$, with $\kappa_c^{\state{uu}}>\pi$. When the vortices become as ``heavy'' as two fluxons, they loose stability because they repel each other and the pinning is very weak. As a result one integer fluxon is emitted and the symmetric FM state turns itself into an asymmetric AFM state discussed above. The value of $\kappa_c$ depends on $a$ and the $\kappa_c(a)$ dependence is presented and discussed below. 

\paragraph{Asymmetric FM state.} The field profiles $\mu_x(x)$ formed at the $(+\kappa,-\kappa)$ discontinuities and corresponding to the asymmetric FM states \state{Uu} are shown in Fig.~\ref{Fig:Vortex2:h}(c) for several values of $\kappa$. This state is stable for any value of $0<\kappa<2\pi$. At $\kappa=0$ or at $\kappa=2\pi$ the state becomes unstable, an integer fluxon is emitted and the state turns itself into a symmetric AFM state. These processes can be written as
\[
  \state{Uu}\stackrel{\kappa=0}{\longrightarrow}\state{du}+\state{F}
  ,\mbox{ or }
  \state{Uu}\stackrel{\kappa=2\pi}{\longrightarrow}\state{UD}+\state{F}.
\]

\paragraph{Symmetric AFM state.} The field profiles $\mu_x(x)$ formed at the $(+\kappa,-\kappa)$ discontinuities and corresponding to the symmetric AFM state \state{du} are shown in Fig.~\ref{Fig:Vortex2:h}(d) for several values of $\kappa$. This state is stable only for $0<\kappa<\kappa_c^{\state{du}}$, $\kappa_c^{\state{du}}>\pi$. When the vortices become ``heavy'', at $\kappa=\kappa_c^{\state{du}}$ they loose stability because they attract each other. They exchange a fluxon (but nothing is emitted!) and the state turns into a complementary one, \ie, $\state{du}\to\state{UD}$. This can be seen in  Fig.~\ref{Fig:Vortex2:h}(d): for $\kappa=1.3\pi$ the system is still in the \state{du} state, but at $\kappa=1.4\pi$ the system is already in the \state{UD} state. Further increase of $\kappa$ results in a decrease of the vortex amplitudes and fluxes as it should be for the \state{UD} state at $\kappa\to 2\pi$. The value of $\kappa_c$ depends on $a$ and the $\kappa_c(a)$ dependence is presented and discussed below.  

%The four possible states and transitions between them are schematically shown in the following diagram
%\begin{CD}
%  sFM @>\kappa=0,2\pi>> a
%\end{CD}

\subsection{Where is the flat phase state?}

%small $a$ limit and correspondence to the 0-pi-LJJ

In the case of $\pi$-vortices the ground state depends on $a$\cite{Goldobin:SF-ReArrange,Kato:1997:QuTunnel0pi0JJ,Zenchuk:2003:AnalXover}. For $a<a_c=\pi/2$ the ground state is the so-called flat phase state $\mu=0$ without magnetic flux. For $a>a_c$ the ground state is made of two AFM ordered semifluxons \state{sa}. In addition there is a FM state \state{ss}, which exists and is stable for any $a$, but its energy is larger than the one of the AFM or flat phase state.

In the case of arbitrary $\kappa$, there is no flat phase state at all, because $\mu=0$, in general, is not a solution of Eq.~(\ref{Eq:sG-mu-static}), with $\theta(x)$ given by Eq.~(\ref{Eq:theta-mm}) or Eq.~(\ref{Eq:theta-pm}). The only exception is $\kappa=\pi n$, with integer $n$.

\begin{figure}[!htb]
  \includegraphics{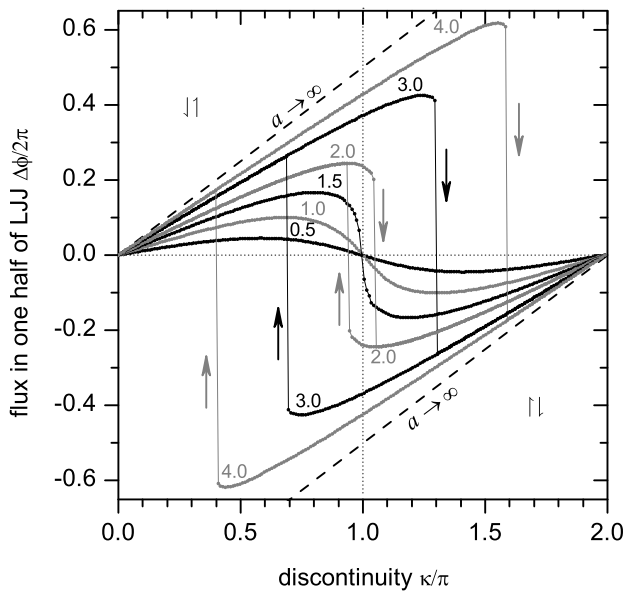}
  \caption{
    The magnetic flux $\Delta\phi$ in the right half of the LJJ ($x>0$) in the symmetric AFM state as a function of $\kappa$ for different values of $a$ shown next to each curve. Dashed line $\Delta\phi=\kappa$ shows the limit of very large $a$.
  }
  \label{Fig:flux-kappa}
\end{figure}

It is therefore interesting to see what happens with the two AFM states discussed above when the distance $a$ between them becomes smaller than $a_c$. We note here that the value of $a_c=\pi/2$ is valid only for $\kappa=\pi$. For other $\kappa$ the value of $a_c$ may change, which is indeed the case for asymmetric AFM state, as we will see below.

\subsubsection{Symmetric AFM state.} 

The crossover distance $a_c^{\state{ud}}=a_c=\pi/2$ is the same and does not depend on $\kappa$. Therefore below we just use $a_c$ to denote it. The phase or magnetic field profiles for the symmetric AFM state look qualitatively the same for $a<a_c$ as well as for $a>a_c$ for any given $\kappa\ne\pi$, see Fig.~\ref{Fig:Vortex2:h}(d).  The only difference is the dependence of the amplitude of magnetic field (or of the flux) of each vortex \vs{} $\kappa$. In Fig.~\ref{Fig:flux-kappa} we show the dependence of magnetic flux
\begin{equation}
  \Delta\phi = \int_{0}^{\infty}\mu_x(x)\,dx=\mu(\infty)-\mu(0)
  %\label{Eq:}
\end{equation}
in one half of the junction (in the whole LJJ the flux integrates to zero for a symmetric AFM state) as a function of $\kappa$ for different distances $a$. In normalized units adopted here one flux quantum is equal to $2\pi$. To plot each curve in Fig.~\ref{Fig:flux-kappa}, we were sweeping $\kappa$ from $0$ to $2\pi$ and back to $0$. The dependence of the vortex's maximum magnetic field on $\kappa$ is qualitatively similar to the behavior of the flux $\Delta\phi(\kappa)$.

As $\kappa$ grows starting from 0, the flux associated which one vortex grows more or less proportional to $\kappa$. If $a$ were very large (not interacting vortices) then the flux of one vortex would be just $\Delta\phi=\kappa$, as shown by dashed lines in Fig.~\ref{Fig:flux-kappa}. For finite $a$, the tails of vortices overlap and the fluxes partially cancel each other resulting in a smaller slope of $\Delta\phi(\kappa)$.

For $a<a_c$, the flux reaches its maximum at some value $0<\kappa_m<\pi$ and starts decreasing (although $\kappa$ increases!), smoothly reaching zero at $\kappa=\pi$. At this point the state of the system is the flat phase state $\mu=0$, which is allowed at $\kappa=\pi$. Further \emph{increase} of $\kappa$ makes the flux \emph{negative}. This corresponds to the smooth continuous transition from the \state{du} to the \state{UD} state as $\kappa$ passes the point $\kappa=\pi$. At $\kappa\to2\pi$ the flux vanishes as it should be for the \state{UD} state. When $\kappa$ is swept back from $2\pi$ to $0$ the flux goes back along the same curve without hysteresis, again making the $\state{UD}\to$ flat phase state $\to\state{du}$ transition at $\kappa=\pi$.

For $a>a_c$ the flux increases almost all the way up to $\kappa_c^{\state{du}}$, passing a maximum just before $\kappa_c^{\state{du}}>\pi$. At $\kappa_c^{\state{ud}}$ the state \state{ud} becomes unstable and abruptly switches to the complementary state \state{UD} as we already saw in Fig.~\ref{Fig:Vortex2:h}(d). At $\kappa\to2\pi$ the flux vanishes as it should be for the \state{UD} state. When $\kappa$ is swept back from $2\pi$ to $0$ the state \state{UD} persists down to $\kappa_c^{\state{UD}}=2\pi-\kappa_c^{\state{du}}$, as it should be for the complementary state, and the system switches back to the original state \state{ud}. 

\begin{figure}[!htb]
  \centering
  \includegraphics{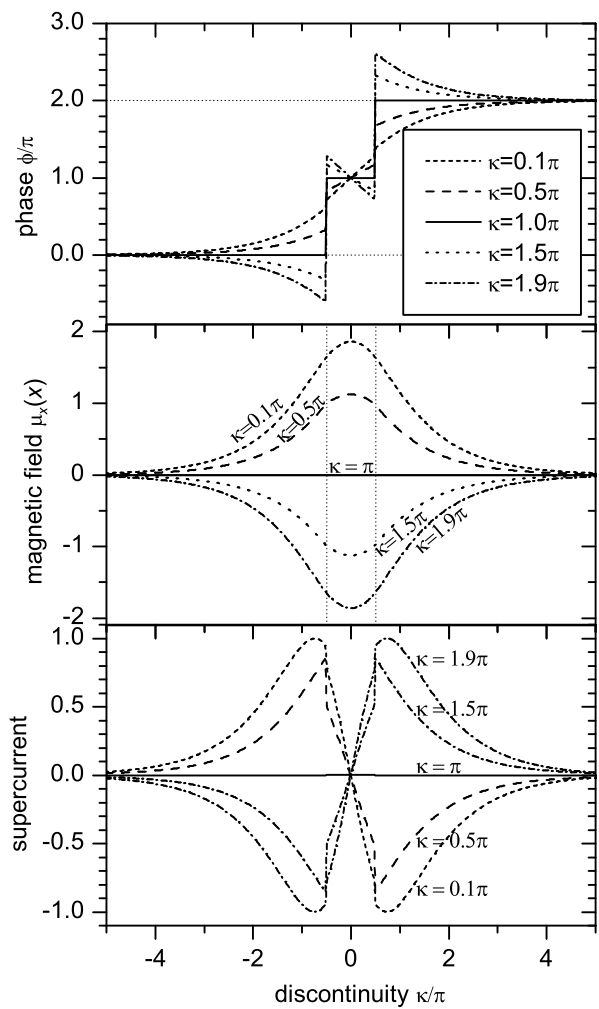}
  \caption{%
    The phase $\phi(x)$, magnetic field $\mu_x(x)$ and the supercurrent of the collective state at $a=1$ for different values of $\kappa$.
  }
  \label{Fig:aAFM:FPS}
\end{figure}
Thus, the crossover which we discovered here is a generalization of the crossover between the AFM state and the flat phase state which was previously investigated for semifluxons\cite{Goldobin:SF-ReArrange,Zenchuk:2003:AnalXover,Kato:1997:QuTunnel0pi0JJ}. The crossover distance $a_c=\pi/2$ is still the same. It is a bifurcation point, such that for $a<a_c$ there exists only one symmetric AFM state (\state{du} or  \state{DU}) for given $\kappa$, while for $a>a_c$ there are two stable symmetric AFM states (\state{du} and \state{DU}) which exist and are stable for $\kappa_c^{\state{UD}}<\kappa<\kappa_c^{\state{du}}$. For $a>a_c$ and $\kappa$ outside this interval, again there is only one stable symmetric AFM state: \state{du} for $0<\kappa<\kappa_c^{\state{UD}}$ or \state{UD} for $\kappa_c^{\state{du}}<\kappa<2\pi$.

\subsubsection{Asymmetric AFM state} 

The phase or magnetic field profiles for the asymmetric AFM state look qualitatively different for $a<a_c^{\state{uD}}$ and for $a>a_c^{\state{uD}}$. Note, that in this case $a_c^{\state{uD}}$ is a weak function of $\kappa$ such that $a_c^{\state{uD}}$ changes from $\pi/2\approx1.57$ at $\kappa=\pi$ towards $a_c^{\state{uD}}\approx1.8$ at $\kappa\to0$ and $\kappa\to2\pi$. In Fig.~\ref{Fig:aAFM:FPS} we show the shape of two vortices in this state for $a=1<a_c^{\state{uD}}$ and different $\kappa$ [\cf, Fig.~\ref{Fig:Vortex2:h}(a) with $a>a_c^{\state{uD}}$]. We see that at $\kappa=\pi$ the state indeed degenerates into a flux-less flat phase state. For $\kappa\ne\pi$, magnetic flux appears, but it is distributed \emph{symmetrically} between the vortices so that $\mu_x(-x)=\mu_x(x)$. In fact, two vortices behave as a single object with the maximum or minimum of the magnetic field at $x=0$. We call this state a \emph{collective} state. When $\kappa\to0$ or $\kappa\to2\pi$ the shape of the vortex approaches the shape of a single fluxon (antifluxon), again centered at $x=0$. In contrast, for $a>a_c^{\state{uD}}$ the asymmetric AFM state at $\kappa\to0$ or $\kappa\to2\pi$ results in a fluxon (antifluxon) centered at $x=\pm a/2$. We do not show the dependence of $\Delta\phi(\kappa)$ similar to the one shown in Fig.~\ref{Fig:flux-kappa} (for the symmetric AFM state) because for the \state{uD} state it is trivial: $\Delta\phi(\kappa)=2\pi-2\kappa$, \ie, it is just a straight line crossing zero at $\kappa=\pi$. The flux above refers to the flux in the \emph{whole} LJJ. We note that this dependence follows from simple topological considerations and is valid for any $a$.

The crossover distance $a_c^{\state{uD}}(\kappa)$ separates the weakly coupled vortices ($a>a_c^{\state{uD}}$) which are in the asymmetric AFM state from the collective strongly coupled vortex state ($a<a_c^{\state{uD}}$) in which the flux is distributed equally and symmetrically between the discontinuities. The dependence $a_c^{\state{uD}}(\kappa)$ can be calculated as an edge of existence of the symmetric solution (solution with maximum $a=a_{\rm max}^{\rm coll}$ for given $\kappa$). The phase plane analysis\cite{Susanto:SF-gamma_c} shows that 
\begin{equation}
  a_c^{\state{uD}}(\kappa) = a_{\rm max}^{\rm coll}(\kappa)
  = 2 \F\left( \frac\pi4 | 1-\sin\frac\kappa2 \right)
  , \label{Eq:Collect:a_c(kappa)}
\end{equation}
is reached when the phase trajectory corresponding to the middle region $|x<a/2|$ of LJJ is tangential to the phase trajectories of the left and the right regions. This happens when the phase in the middle region changes from $\mu(-a/2)=(\pi-\kappa)/2$ to $\mu(+a/2)=3(\pi-\kappa)/2$. The function $\F(\varphi|m)$ in Eq.~(\ref{Eq:Collect:a_c(kappa)}) is the \emph{incomplete elliptic integral of the first kind}\cite{Abramowitz-Stegun}.Note, the limiting behavior $a_c^{\state{uD}}(\pi)=\pi/2$, while $a_c^{\state{uD}}(0)=a_c^{\state{uD}}(2\pi)=2\ln(1+\sqrt2)\approx1.7628$, which is in good agreement with numerical results.

\subsection{What happens at $\kappa_c$?}

\begin{figure*}[!htb]
  \begin{center}
   \includegraphics{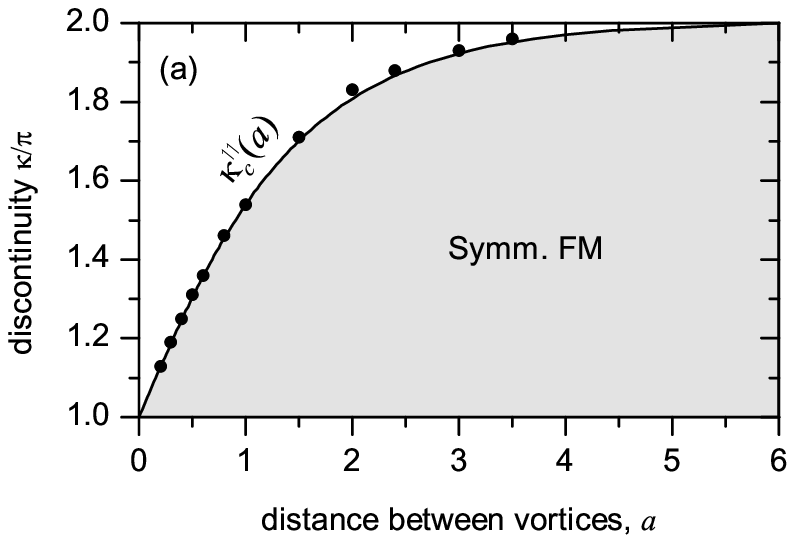}
   \hfill\nobreak
   \includegraphics{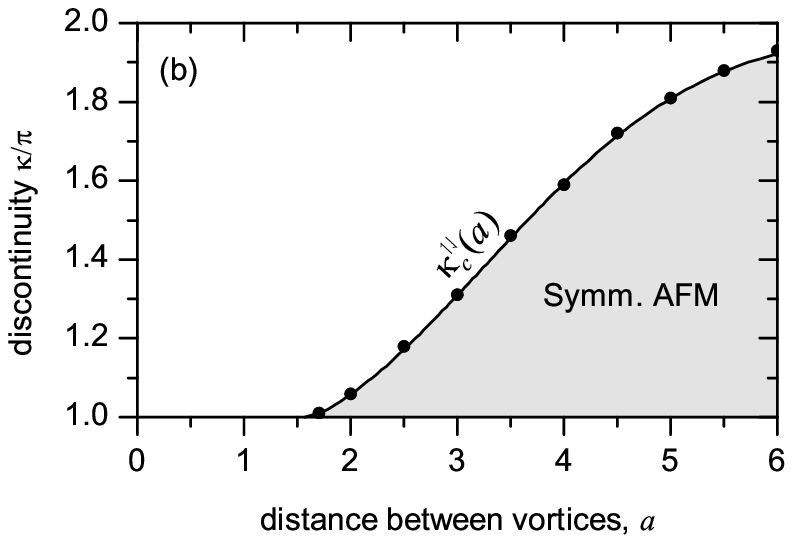}
  \end{center}
  \caption{%
    The behavior of $\kappa_c(a)$ for symmetric FM (a) and AFM (b) states of two vortices calculated using direct numerical simulation of Eq.~(\ref{Eq:sG-mu}) shown by symbols. The continuous lines show the boundaries of existence regions $a_s^{\protect\state{uu}}(\kappa)$ given by Eq.~(\ref{Eq:sFM:a_c(kappa)}) (a) and $a_s^{\protect\state{ud}}(\kappa)$ (b) calculated using the phase plane analysis. The area of existence \emph{and} stability of each state is shadowed.  
  }
  \label{Fig:kappa_c(a)}
\end{figure*}

The values of $\kappa_c^{\state{du}}(a)$ and $\kappa_c^{\state{uu}}(a)$ are  calculated numerically and are shown as symbols in Fig.~\ref{Fig:kappa_c(a)}. The boundary of existence for each solution can be derived using a phase plane analysis\cite{Susanto:SF-gamma_c}. In fact, instead of searching for the critical value of $\kappa$ at given $a$, in the phase plane analysis we search for the switching distances $a_s^{\state{ud}}$ and $a_s^{\state{uu}}$ as a function of the discontinuity $\kappa$. 

First, we treat symmetric FM state. At the edge of the solution existence the phase trajectories 
\begin{eqnarray}
  \mu_x^{(0)}(\mu)&=&\pm\sqrt{1-\cos(\mu)}
  ; \label{Eq:SymmFM:PhaseTraj1}\\
  \mu_x^{(\kappa)}(\mu)&=&\pm\sqrt{C-\cos(\mu-\kappa)}
  ; \label{Eq:SymmFM:PhaseTraj1}\\
  \mu_x^{(2\kappa)}(\mu)&=&\pm\sqrt{1-\cos(\mu-2\kappa)}
  , \label{Eq:SymmFM:PhaseTraj1}
\end{eqnarray}
corresponding to the 0, $\kappa$, and $2\kappa$ regions, touch each other. This happens at value of integration constant $C=1+2\sin(\kappa/2)$, and the phase $\mu$ in the middle $\kappa$-region changes from $(\kappa+\pi)/2$ at $x=-a/2$ to $\kappa$ at $x=0$ and to $(3\kappa-\pi)/2$. At this state the value of $a$ can be calculated as
\begin{equation}
  a_s^{\state{uu}}(\kappa) %=
  = 2\sqrt{m_{\rm FM}}\left[
  \F\left( \frac\pi2 | m_{\rm FM}\right) - 
  \F\left( \frac{3\pi-\kappa}{4} | m_{\rm FM} \right)
  \right]
  , \label{Eq:sFM:a_c(kappa)}
\end{equation}
with modulus
\[
  m_{\rm FM}(\kappa) = \frac{1}{1+\sin\frac\kappa2}.
\]
Note, that $\F(\pi/2|m)\equiv\K(m)$ which is a \emph{complete elliptic integral of the first kind}\cite{Abramowitz-Stegun}. The edge of solution existence given by Eq.~(\ref{Eq:sFM:a_c(kappa)}) is plotted in Fig.~\ref{Fig:kappa_c(a)}(a) as continuous line. One can see the perfect agreement between Eq.~(\ref{Eq:sFM:a_c(kappa)}) and the result of direct numerical simulation of Eq.~(\ref{Eq:sG-mu}). At $a\to0$, $\kappa_c^{\state{uu}}(a)\approx\pi+2a$. This limiting behavior is clear, as two semifluxons with $a=0$ form a fluxon. As soon as we try to increase $\kappa$ the state becomes unstable, because a vortex with the topological charge larger than $2\pi$ is unstable\cite{Goldobin:2KappaEigenModes}.

Second, we analyze the symmetric AFM state. If one draws the trajectories 
\begin{eqnarray}
  \mu_x^{(0)}(\mu)&=&\pm\sqrt{1-\cos(\mu)}
  ; \label{Eq:SymmAFM:PhaseTraj1}\\
  \mu_x^{(\kappa)}(\mu)&=&\pm\sqrt{C-\cos(\mu-\kappa)}
  ; \label{Eq:SymmAFM:PhaseTraj1}\\
  \mu_x^{(2\kappa)}(\mu)&=&\pm\sqrt{1-\cos(\mu)}
  , \label{Eq:SymmAFM:PhaseTraj1}
\end{eqnarray}
corresponding to the 0-$\kappa$-0 LJJ on the phase plane, one can see that the symmetric AFM solution exists when the 
integration constant $C$ changes from $C=1$ (corresponding to $a=\infty$) down to $C_c=1-2\sin(\kappa/2)$ (some finite $a$) corresponding to phase trajectories for 0 and $\pi$ regions touching each other. The naive conclusion that the state with $C=C_c$ will correspond to the switching (minimum) distance $a_s$ is wrong! As it was pointed out by Susanto\cite{Sustanto:2004:PrivComm}, the dependence $a(C)$ is not monotonous and has a minimum on the interval $C=C_c(\kappa) \ldots 1$. It is this minimum distance, which corresponds to $a_s^{\state{ud}}$. 

For given $\kappa$ and $C$ the distance $a$ is given by
\begin{equation}
  a(C,\kappa)= 2\left[ 
  \K\left( \frac{1+C}{2} \right) - 
  \F\left( \varphi_0(C,\kappa) | \frac{1+C}{2} \right)
  \right]
  , %\label{Eq:}
\end{equation}
where\cite{Sustanto:2004:PrivComm}
\[
  \varphi_0(C,\kappa)=\sgn(\theta_0(C,\kappa))\arcsin\sqrt\frac{1-\cos\theta_0(C,\kappa)}{1+C},
\]
and
\[
  \theta_0(C,\kappa) = \pi-\frac{\kappa}{2}
  -\arcsin\frac{1-C}{2\sin\frac{\kappa}{2}}
\]
The value of $C$ at which $a(C,\kappa)$ has a minimum for given $\kappa$ can be found from $\partial a(C,\kappa)/\partial C=0$. One ends up with a rather bulky transcendental equation for $C$. For the sake of simplicity we find the minimum value $a_{\rm min}(\kappa)=a_s^{\state{ud}}(\kappa)$ for given $\kappa$ numerically. The result is shown in Fig.~\ref{Fig:kappa_c(a)}(b) by a continuous line. One can see the perfect agreement between the result obtained from the phase plane analysis and the result of direct numerical simulation of Eq.~(\ref{Eq:sG-mu}). Note, that $\kappa_c^{\state{ud}}\to\pi$ when $a \to a_c=\pi/2$.

At large $a\to\infty$ both symmetric AFM and FM states become unstable when each vortex grows up to a fluxon size ($\kappa=2\pi$). This is a natural limit for non-interacting vortices.

\subsection{Lowest energy state}

One may ask, what is the lowest energy state among the four irreducible states discussed above? As we know, for a semifluxon the AFM state (for $a>a_c$) or the flat phase state (for $a<a_c$) \emph{always} has a lower energy than the FM state\cite{Goldobin:SF-ReArrange}, \ie, for any $a$. Is this the case for arbitrary vortices, too? Before answering this question, we note that the competition of energies can only be considered among the states with the same structure of discontinuities. 

\begin{figure*}[!tb]
  \begin{center}
    \includegraphics{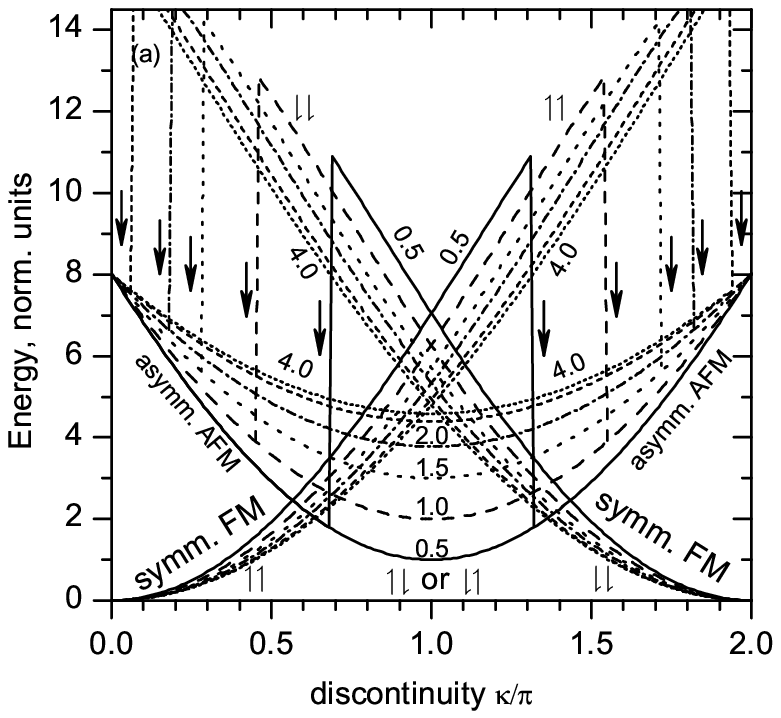}
    \hfill\nobreak
    \includegraphics{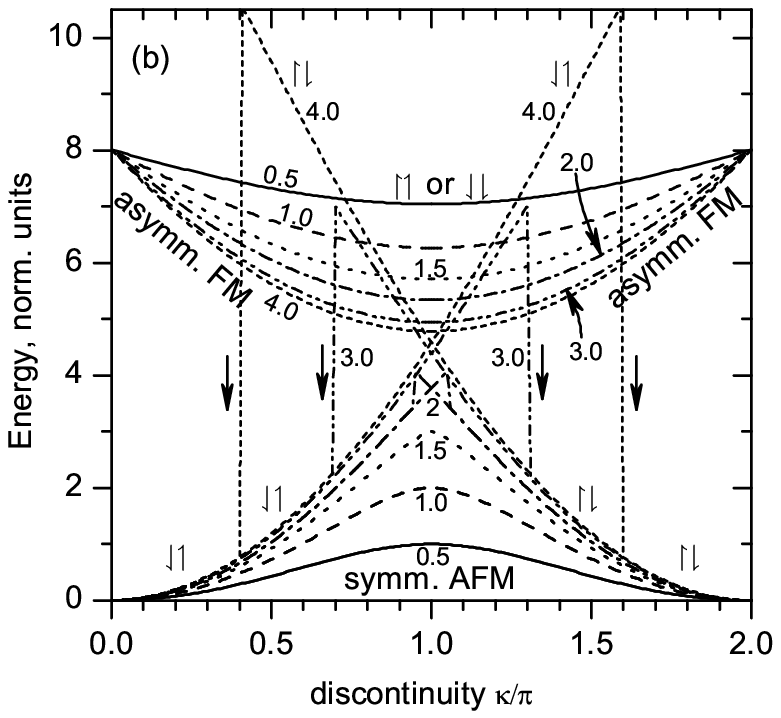}
  \end{center}
  \caption{%
    Numerically calculated energy $E(\kappa)$ of competing states for different values of $a$, shown next to each curve. (a) asymmetric AFM \vs{} symmetric FM; (b) symmetric AFM \vs{} asymmetric FM;
  }
  \label{Fig:E-kappa}
\end{figure*}

Thus the first question is: given two discontinuities $(-\kappa, -\kappa)$ at a distance $a$ between them, which state has lower energy: the asymmetric AFM or the symmetric FM? Looking at Fig.~\ref{Fig:Vortex2:h}(a)--(b), one can qualitatively say, that at small $\kappa$ the symmetric FM state has lower energy than the asymmetric AFM state. Instead, at large $\kappa$ the AFM state should have lower energy. The numerically calculated energy $E(\kappa)$ of both states is plotted in Fig.~\ref{Fig:E-kappa}(a) for different values of $a$. To calculate the energy we have solved numerically Eq.~(\ref{Eq:sG-mu}) to get the solution $\mu(x)$ and calculated the energy by integrating Eq.~(\ref{Eq:HamiltonianDensity}) over the junction length. As one can see in Fig.~\ref{Fig:E-kappa}(a), the asymmetric AFM state has lower energy than \emph{both} symmetric FM states only in a rather narrow interval of $\kappa$ around $\kappa=\pi$. If $\kappa$ is outside of this interval one of the symmetric FM states has lower and another higher energy than the corresponding asymmetric AFM state at the same $\kappa$. In Fig.~\ref{Fig:E-kappa}(a) one can also see transitions at $\kappa_c$ corresponding to the switching from the ``heavy'' symmetric FM state to the asymmetric AFM state.

The second question is: given two discontinuities $(+\kappa, -\kappa)$ and the distance $a$ between them, which state has lower energy, the symmetric AFM or the asymmetric FM? Looking at Fig.~\ref{Fig:Vortex2:h}(c)--(d), one can qualitatively say, that at small $\kappa$ the symmetric AFM state has lower energy. At larger $\kappa$ it is not so clear. The calculations give the $E(\kappa)$ plots shown in Fig.~\ref{Fig:E-kappa}(b). As we see, in the absence of hysteresis ($a<a_c$) the symmetric AFM state always has lower energy than the asymmetric FM state. For large $a$, the most ``heavy'' of the symmetric AFM states may have larger energy than the corresponding asymmetric FM state at the same $\kappa$.

\section{Conclusions}
\label{Sec:Conclusion}

We investigated possible ground states of fractional vortices formed at one and two $\kappa$-discontinuities of the phase.

In case of one $-\kappa$-discontinuity we derived the shape of a \emph{direct} $+\kappa$-vortex, see Eqs.~(\ref{Eq:dvortex}) and (\ref{Eq:dvortex:x_0}), and a \emph{complementary} $(\kappa-2\pi)$-vortex, see Eqs.~(\ref{Eq:cvortex}) and (\ref{Eq:cvortex:x_0}), as well as their energy as a function of $\kappa$ (\ref{Eq:E-kappa}). In the general case $\kappa\ne\pi$ these vortices are \emph{not mirror symmetric} like semifluxons.

\begin{figure*}[!htb]
  \begin{center}
    \includegraphics{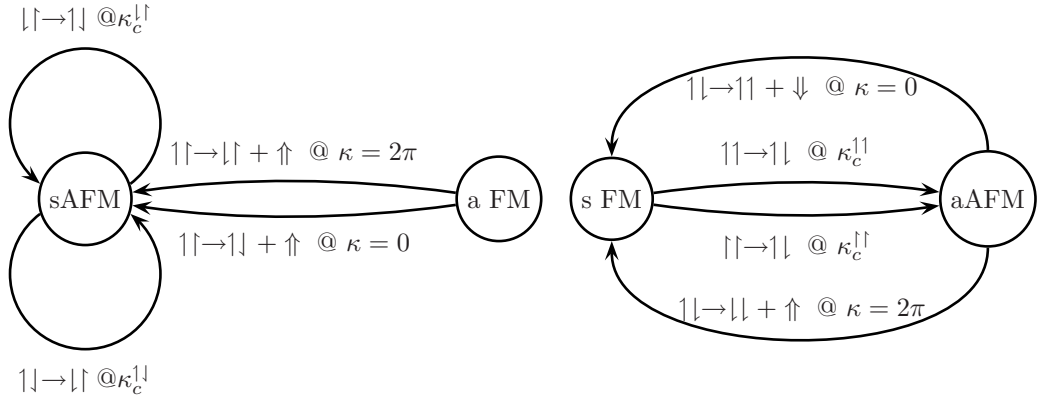}
  \end{center}
  \caption{%
    Possible transitions between different weakly coupled two vortex states which can be achieved by varying $\kappa$.
  }
  \label{Fig:state-diagram}
\end{figure*}

Due to such a broken symmetry, the ground state of the system with two discontinuities consists of four different states: asymmetric and symmetric AFM and FM states. Symmetric states consist of two direct vortices or two complementary vortices, while asymmetric states consist of one direct and one complementary vortex. Asymmetric states are stable for $0<\kappa<2\pi$ and their properties are symmetric (energy is the same, magnetic field just changes sign) with respect to the transformation $\kappa\to2\pi-\kappa$. Symmetric states exist only in some interval of $\kappa$. For example, the state \state{du} exists only for $0<\kappa<\kappa_c^{\state{du}}\geq\pi$ and the state \state{uu} exists only for $0<\kappa<\kappa_c^{\state{uu}}\geq\pi$. The values $\kappa_c^{\state{du}}(a)$ and $\kappa_c^{\state{uu}}(a)$ have been found, see Fig.~\ref{Fig:kappa_c(a)}. If $\kappa$ exceeds $\kappa_c$, the state turns itself into another state, \eg, $\state{ud}\to\state{DU}$ or $\state{uu}\to\state{uD}$. For the symmetric AFM state a new, more general, meaning of the crossover distance $a_c^{\state{ud}}=a_c=\pi/2$ between the discontinuity points is discovered. If $a<a_c^{\state{ud}}$, the transition between the \state{ud} and \state{DU} states is smooth, otherwise it is abrupt and is associated with the instability of the \state{ud} state at $\kappa=\kappa_c^{\state{du}}$. The transitions between different weakly coupled states which can be induced by changing $\kappa$ are summarized in Fig.~\ref{Fig:state-diagram}.

The asymmetric AFM state exists only for $a>a_c^{\state{uD}}(\kappa)$, while for $a<a_c^{\state{uD}}(\kappa)$ it turns itself into a new strongly coupled ``collective'' state which is perfectly symmetric and can not be simply constructed out of the single vortex states. The crossover distance $a_c^{\state{uD}}$ is a weak function of $\kappa$ given by Eq.~(\ref{Eq:Collect:a_c(kappa)}). 

Finally, we have calculated and compared the energies of competing states. We showed that the energy of the asymmetric AFM state can be larger as well as smaller than the one of the symmetric FM state depending on $\kappa$. For fixed $\kappa$ it may or may not depend on $a$. On competition between symmetric AFM and asymmetric FM states, there is always (for any $a$ and $\kappa$) a symmetric AFM state which has lower energy than the asymmetric FM state. The details are presented in Fig.~\ref{Fig:E-kappa}.

As we see, the variety of the ground states of arbitrary vortices is much more rich than the one of semifluxons. The asymmetry may be exploited in information processing devices based on distinction and controllable manipulation of the vortex states.

The case of two vortices represents the simplest system of two coupled vortices. Of course, in the case of $N>2$ discontinuities, one should expect even a larger variety of possible vortex states which may strongly depend on the parity of $N$, but the treatment of those configurations should depend on practical needs. In this sense the case of two vortices is especially important as it is the first candidate for implementation of classical or quantum bits based on fractional vortices. The difficulty with one vortex is that it emits a fluxon when it flips. The AFM state, instead can flip like $\state{ud}\leftrightarrow\state{du}$ or $\state{dU}\leftrightarrow\state{Ud}$ without any emission.

\begin{acknowledgments}
  We acknowledge fruitful discussions with H.~Susanto. This work was supported by the Deutsche Forschungsgemeinschaft project GO-1106/1, and by the ESF programs "Vortex" and "Pi-shift".
\end{acknowledgments}

\bibliography{LJJ,pi,SF,QuComp,SFS,software,this}

\end{document}